\documentclass[journal]{IEEEtran}
\IEEEoverridecommandlockouts
\usepackage{cite}

\usepackage{amsmath,amssymb,amsfonts,mathtools,bm}
\usepackage{amsthm}
\usepackage{algorithm}
\usepackage{algorithmic}
\usepackage{graphicx}
\usepackage{textcomp}
\usepackage{xcolor,float}
\usepackage{tabularx}
\usepackage{textcomp}
\usepackage{diagbox}
\usepackage{svg}
\usepackage{booktabs}
\usepackage{subfigure}
\usepackage{hyperref}
\usepackage{graphicx}
\usepackage[margin=0.5in]{geometry}
\usepackage{changes}

\newtheorem{problem}{Problem}
\newtheorem{lemma}{Lemma}

\usepackage{tikz,xcolor,hyperref}% Make Orcid icon
\definecolor{lime}{HTML}{A6CE39}
\DeclareRobustCommand{\orcidicon}{%
    \begin{tikzpicture}
    \draw[lime, fill=lime] (0,0) 
    circle [radius=0.16] 
    node[white] {{\fontfamily{qag}\selectfont \tiny ID}};    \draw[white, fill=white] (-0.0625,0.095) 
    circle [radius=0.007];    \end{tikzpicture}
    \hspace{-2mm}}
\foreach \x in {A, ..., Z}{%
    \expandafter\xdef\csname orcid\x\endcsname{\noexpand\href{https://orcid.org/\csname orcidauthor\x\endcsname}{\noexpand\orcidicon}}
    }
\allowdisplaybreaks
\renewcommand{\baselinestretch}{1}

\begin{document}

% \title{Effective Multi-UAV Trajectory Design for AoI Management: A GNN Empowered Partial Observation MARL Method}
\title{GNN-Empowered Effective Partial Observation MARL Method for AoI Management in Multi-UAV Network}

\author{{Yuhao Pan\orcidA{},}
Xiucheng Wang\orcidB{},~\IEEEmembership{Graduate Student Member,~IEEE,}
% Jie Gao\orcidC{},~\IEEEmembership{Senior Member,~IEEE,}
Zhiyao Xu\orcidD{},
Nan Cheng\orcidE{},~\IEEEmembership{Senior Member,~IEEE,}
Wenchao Xu\orcidF{},~\IEEEmembership{Member,~IEEE,}
Jun-jie Zhang\orcidG{}
% <-this % stops a space
% 左下角的致谢
\thanks{ }% <-this % stops a space
\thanks{
\par This work was supported by the National Key Research and Development Program of China (2020YFB1807700), and the National Natural Science Foundation of China (NSFC) under Grant No. 62071356.
\par Yuhao Pan is with the School of Electronic Engineering, Xidian University, Xi'an 710071, China (e-mail:yhpan@stu.xidian.edu.cn). \textit{Yuhao Pan and Xiucheng Wang contribute equally.} 
\par Xiucheng Wang and Nan cheng are with the State Key Laboratory of ISN and School of Telecommunications Engineering, Xidian University, Xi’an 710071, China (e-mail:xcwang\_1@stu.xidian.edu.cn; dr.nan.cheng@ieee.org). \textit{Nan Cheng is the corresponding author.}
\par Zhiyao Xu is with the School of Artificial Intelligence, Xidian University, Xi'an, 710071, China (e-mail:21009200843@stu.xidian.edu.cn).
\par Wenchao Xu is with the Department of Computing, The Hong
Kong Polytechnic University, Hong Kong, China (e-mail: wenchao.xu@polyu.edu.hk).
\par Jun-jie Zhang is with the Northwest Institute of Nuclear Technology, Xi’an 710024, China (e-mail: zhangjunjie@nint.ac.cn).
}
}

    \maketitle

\IEEEdisplaynontitleabstractindextext

\IEEEpeerreviewmaketitle

\begin{abstract}
Unmanned Aerial Vehicles (UAVs), due to their low cost and high flexibility, have been widely used in various scenarios to enhance network performance. However, the optimization of UAV trajectories in unknown areas or areas without sufficient prior information, still faces challenges related to poor planning performance and low distributed execution. These challenges arise when UAVs rely solely on their own observation information and the information from other UAVs within their communicable range, without access to global information. To address these challenges, this paper proposes the Qedgix framework, which combines graph neural networks (GNNs) and the QMIX algorithm to achieve distributed optimization of the Age of Information (AoI) for users in unknown scenarios. The framework utilizes GNNs to extract information from UAVs, users within the observable range, and other UAVs within the communicable range, thereby enabling effective UAV trajectory planning. Due to the discretization and temporal features of AoI indicators, the Qedgix framework employs QMIX to optimize distributed partially observable Markov decision processes (Dec-POMDP) based on centralized training and distributed execution (CTDE) with respect to mean AoI values of users. By modeling the UAV network optimization problem in terms of AoI and applying the Kolmogorov-Arnold representation theorem, the Qedgix framework achieves efficient neural network training through parameter sharing based on permutation invariance. Simulation results demonstrate that the proposed algorithm significantly improves convergence speed while reducing the mean AoI values of users. The code is available at \url{https://github.com/UNIC-Lab/Qedgix}.
\end{abstract}


\section{Introduction}
In the landscape of communication networks, the Internet of Things (IoT) has emerged as an important domain \cite{akpakwu2017survey}. IoT excels in monitoring environmental variables and facilitates the widespread deployment of sensors and devices. It is particularly effective at handling small-sized data transmissions and can also accommodate applications with varying degrees of time sensitivity \cite{kang2022personalized,hazra2023fog,bian2022machine,kang2023blockchain}, such as real-time environmental monitoring and smart metering. Fresh data can enhance the decision-making capabilities of the central processing units within IoT systems \cite{acharya2018big}, rendering the Age of Information (AoI) an important metric for data collection in IoT \cite{abd2019role,zhang2023aoi}. In an ideal scenario, devices should regularly and continuously upload data to minimize the AoI. However, in remote areas, the continuous collection of device data can be challenging both economically and operationally \cite{yaacoub2020key, wang2024reliable}. Unmanned Aerial Vehicles (UAVs), serving as dynamic network access nodes, provide a flexible and cost-effective solution \cite{motlagh2017uav,bai2020air,dai2022unmanned,bai2023towards,li2020energy}. The deployment and trajectory optimization of UAVs have become focal points in recent research. Reinforcement Learning (RL) is commonly employed as a method for optimizing these aspects over time. Utilizing RL helps to efficiently manage the movement and operational decisions of UAVs, adapting to changing conditions and requirements. Although RL can be applied to UAVs trajectory planning for minimizing the AoI, it often results in suboptimal performance due to significant challenges.

Two major challenges in leveraging RL for optimizing trajectories to minimize the AoI in IoT systems are the efficiency of training and the performance of the output trajectories. The first challenge, training efficiency, is crucial because RL models require substantial data and computational resources, which can be limited in real-world scenarios \cite{ibarz2021train}. The second challenge, the performance of the optimized trajectories, is to ensure that the trajectories generated by the models are practical and reliable in varying environmental conditions. These challenges reflect the ongoing struggle to adapt advanced UAV technologies within the practical constraints of IoT system deployments \cite{fotouhi2019survey, cheng2023ai,xu2024blockchain}. In such scenarios, employing centralized RL with a single agent presents challenges, especially as the action space for trajectory planning expands exponentially with the increase in the number of users \cite{aradi2020survey}. This single agent, tasked with determining the optimal flight paths based on the geographic distribution of user demands, must continuously make choices from a vast array of potential UAV flight actions. Consequently, this often leads to inadequate exploration of the action space, resulting in substantial convergence difficulties in identifying optimal actions. Some researchers adopt multi-agent reinforcement learning (MARL) algorithms, in which each agent is assigned to manage the trajectory planning for a specific UAV, drastically narrowing the action space for that particular UAV. This shift not only simplifies the decision-making process but also enhances the overall efficiency and effectiveness of the system. 

However, existing MARL algorithms, such as the policy gradient-based Multi-Agent Deep Deterministic Policy Gradient (MADDPG) or the value-based QMIX \cite{lowe2017multi}, have limitations. Specifically, these algorithms usually assume that each agent has access to the global state to achieve optimal performance, a premise often impractical in real-world applications. When operating MARL algorithms in tasks aimed at optimizing AoI, agents, in this context the UAVs, need to have as much global information as possible about the state of all UAVs and users. Due to the limited range of their sensing capabilities, UAVs are unable to obtain global state information \cite{wang2017autonomous}. Therefore, it is imperative to make critical adjustments to MARL algorithms. These include enabling the algorithms to operate with only partial state information and to effectively carry out real-time UAV trajectory planning tasks. These adjustments are essential for maintaining a balance between comprehensive state awareness and efficient trajectory optimization.

In the field of multi-agent systems for wireless communications, collaboration between agents is crucial for optimizing overall system performance \cite{xiao2022perception,feriani2021single,xiao2023overcoming,yu2022fully,gong2011joint}. Enhancing agents' perception capabilities through exchanging information with other agents allows each one to have a more comprehensive understanding of the global state, thereby enabling better collaboration and task allocation. A key challenge is identifying the optimal frequency and scope of these exchanges, which can be elusive in the preliminary analysis of system optimization. Interestingly, this challenge shares similarities with the message-passing mechanisms used in graph neural networks (GNNs). In both cases, the goal is to efficiently extract and utilize relevant features through the exchange of information \cite{wang2023scalable}. In GNNs, message passing aggregates features from neighboring nodes to improve the representation of each node, which is similar to how agents in a multi-agent system exchange information to enhance their perception and coordination. In this process, each agent, acting as a GNN node, transmits messages to others, aiding feature extraction from a wider range of subgraphs formed by neighboring nodes \cite{kortvelesy2022qgnn}, compared to relying only on self-observation. Empirical evidence from GNNs applications shows that extracting features from one or two-hop neighbor nodes can often lead to satisfactory results, typically requiring only 1 to 2 GNN layers \cite{xu2018powerful}. Considering these insights, our method treats each agent as a GNN node. Initially, each agent extracts local information based on its observations. Then, each agent shares these features with neighboring agents in the graph and combines the received neighboring features with its own observations to enhance decision-making.

The use of GNNs for decision-making in multi-agent systems raises an important question: can GNNs directly utilize the reward from the environment feedback loop to achieve multi-agent optimization? The conclusion is negative. Relying solely on the node feature output by GNNs to implement multi-agent systems is analogous to using only the policy network of each agent in the traditional MARL framework, thereby neglecting the effective gradient information from the critic network or the mixer network for training. To achieve efficient GNN training, we integrate the QMIX framework, treating the GNN as a policy network that effectively extracts environmental features and makes decisions. We train the GNN based on the global reward, specifically the mean AoI values of the users, by cascading its output into the mixer network. This cascade framework inherently brings about output permutation invariance. Through permutation invariant analysis of the optimization problem, we realize an efficient training method based on parameter sharing by employing the Kolmogorov-Arnold representation theorem. Our contributions in this paper are
summarized as follows.
\begin{enumerate}
    \item The distributed UAV trajectory planning problem, aimed at optimizing mean AoI values without global information, is modeled as a Distributed Partially Observable Markov Decision Process (Dec-POMDP). We demonstrate that user-average AoI is permutation-invariant with respect to UAV locations.
    \item Leveraging the permutation invariance of the optimization problem and the Kolmogorov-Arnold representation theorem, we propose an efficient distributed AoI optimization using a cascade architecture of GNN and QMIX with permutation invariant properties. GNNs are utilized to extract features of UAVs and users and to optimize UAV movement during distributed execution. The QMIX network translates the discrete time series index of mean AoI values of users into an effective gradient that updates the GNN parameters, which is then omitted during execution to facilitate distributed optimization.
    \item Extensive simulation experiments demonstrate that the proposed method exhibits superior performance and faster convergence speed compared to methods using QMIX alone or a combination of QMIX and GNN.
\end{enumerate}

\section{System Model and Problem Formulation}
\subsection{System Model}
\begin{figure}[ht]
  \centering
  \includegraphics[width=0.95\columnwidth]{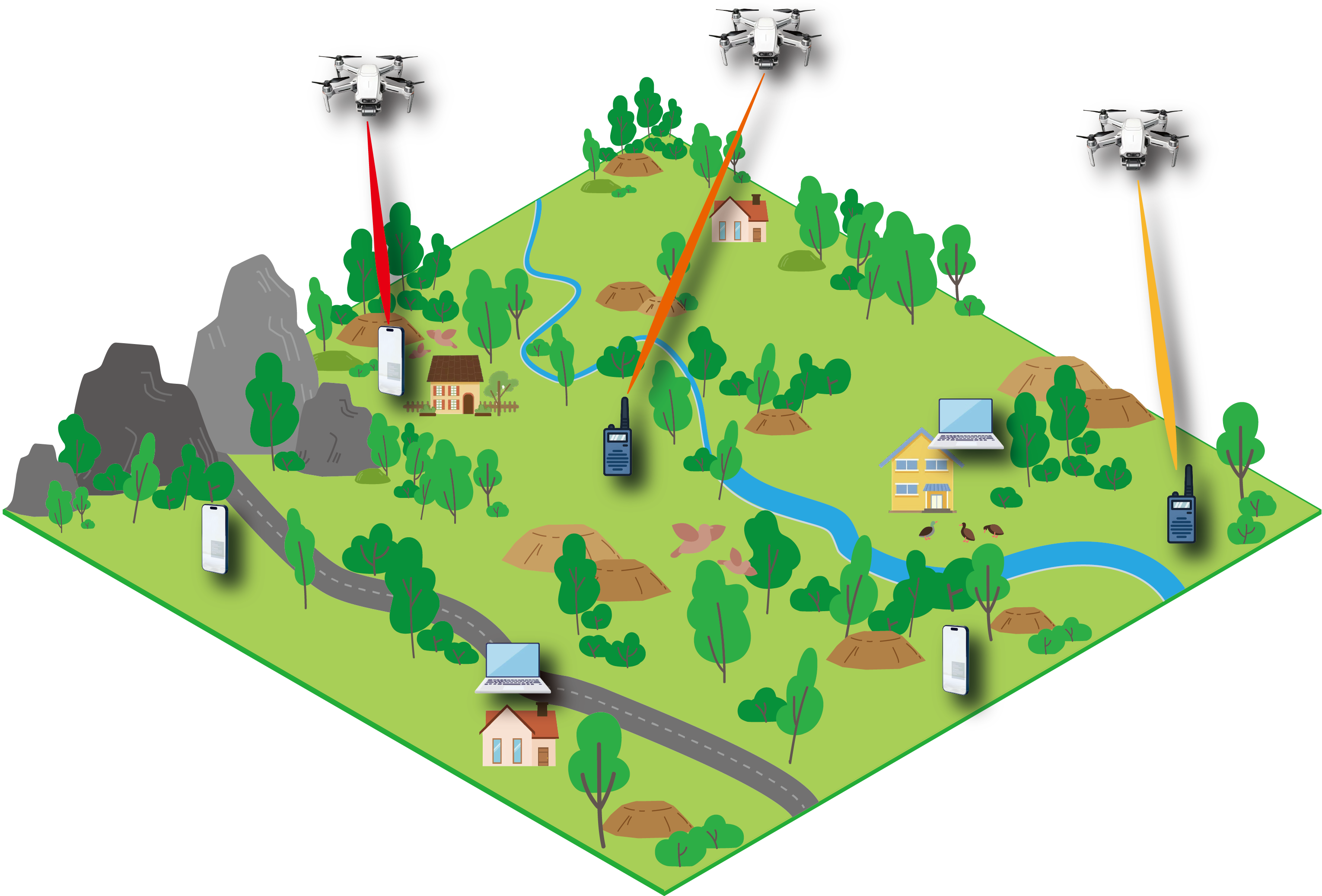}
   % \vspace{-9pt}
  \centering \caption{The UAVs collect data from users in remote areas.} 
  \label{background}
   % \vspace{-9pt}
\end{figure}
Consider the following scenario. $N$ IoT users are distributed in a remote area without ground communication infrastructure. $M$ UAVs are used to collect data about these users. In a practical scenario, $M \leq N$, and the overall coverage of all UAVs at any instant is smaller than the considered area, as in \cite{wang2022joint}, it is impossible to collect data of all users by optimizing the deployment location of UAVs. Therefore, in order to collect the latest data from all users as much as possible, it is necessary to optimize the trajectories of the UAVs. Similar to \cite{zhou2019deep}, to emphasize the reusability of UAV trajectory planning for optimizing transmission coverage for data collection, transmission rate and data packet size are not considered. Thus, as long as a user is in the transmission range of a UAV, the UAV can collect the user's data instantly. In the environment, all users remain stationary, and the coordinates of user $i$ can be represented by the tuple $\left(x_{i}^{user},y_{i}^{user}\right)$. In time slot $k$, the coordinates of UAV $i$ can be represented by the tuple $\left(x_{k,i}^{uav},y_{k,i}^{uav}\right)$.

\begin{table}[ht]
    \centering
    \caption{Important Notations Used in This Paper.}
    % 调整行高
    \renewcommand{\arraystretch}{1.15}
    \resizebox{0.95\linewidth}{!}{\begin{tabular}{c|c}
    \toprule[0.5pt]\toprule[0.5pt]
    \textbf{Notation} & \textbf{Definition} \\ \hline
    $k$ &  Index of time slot. \\
    $M$ & Number of UAVs. \\
    $N$ & Number of IoT users. \\
    $d$ & Detection range. \\
    $t$ & Transmission range. \\
    $\mathcal{C}_{k}$ & Transmission coverage area in slot $k$. \\
    % $\mathcal{D}_{k,j}$ & Detection range area of UAV $j$ in slot $k$. \\
    $a_{i}^{k}$ & AoI for user $i$ in slot $k$. \\
    $x_{i}^{user},y_{i}^{user}$ & Location of user $i$. \\
    $x_{k,j}^{uav},y_{k,j}^{uav}$ & Location of UAV $j$ in slot $k$. \\
    $Q_{tot}^{k}$ & The estimated global Q-value in slot $k$. \\
    $\vartheta_{k}^{j}$ & Flight direction of UAV $j$ in slot $k$. \\
    $o_{i}$ & Observation of agent $i$. \\
    $r_k$ & Instantaneous reward in slot $k$. \\
    $v$ & UAV speed. \\
    $\gamma$ & Discount factor. \\
    $\mathbf{A}$ &  Graph adjacency matrix. \\
    $\mathbf{O}^{v}_{(i,:,:)}$ &  UAVs' coordinates observed by the $i$-th node. \\
    $\mathbf{O}^{u}_{(i,:,:)}$ &  Users' information observed by the $i$-th node. \\
    $\xi$ & Length measurement unit in experiments. \\
    \bottomrule[0.5pt]\bottomrule[0.5pt]
    \end{tabular}}
    \label{notations}
\end{table}

The UAVs have two ranges: the detection range, denoted as $d$, and the transmission range, denoted as $t$, where $d \geq t$ \cite{zeng2016wireless}. The UAV's detection range is a circular region with a radius of $d$, employing sensors to detect nearby entities such as users and other UAVs and acquiring their positional information, including the AoI if a user is detected. The set $\mathcal{D}_{k,j}$ denotes the detection range area of UAV $j$ in time slot $k$. The transmission range of the UAV is a circular area with a radius of $t$. A user $i$ falls within the transmission range of UAV $j$ in time slot $k$ if $(x_{i}^{user}-x_{k,j}^{uav})^2+(y_{i}^{user}-y_{k,j}^{uav})^2 \leq t^2$, which enables the UAV to instantly collect data from user $i$. The set $\mathcal{C}_{k,j}$ represents the transmission range area of UAV $j$ in time slot $k$. The joint transmission range area set $\mathcal{C}_{k}$ of all UAVs is defined as $\mathcal{C}_{k}=\{\mathcal{C}_{k,1}\cup\cdots\cup\mathcal{C}_{k,M}\}$.

The AoI, defined as the time elapsed since the user last successfully transmitted a message, is adopted as the metric for data collection. The AoI for user $i$ in time slot $k$ is:
\begin{align}
    a_{i}^{k} =
    \begin{cases}
        a_{i}^{k} + 1, & \text {$(x_{i}^{user},y_{i}^{user})\notin \mathcal{C}_{k}$},  \\
        0, & \text {$(x_{i}^{user},y_{i}^{user})\in \mathcal{C}_{k}$},
    \end{cases}
\end{align}
% where the above formula represents the situation where the AoI value changes with the time stamp. When $k=0$, the AoI is 0. As the time index increases, if the user's data remains uncollected, the AoI value will increase. However, if the user is within the coverage of the UAV, the data is uploaded, and the AoI value is reset to zero. This indicator is of significant importance. When AoI is high, there is a priority where By adopting AoI as the metric UAVs are encouraged to quickly proceed to the location of users with higher AoI values in order to collect information promptly. Through research, it is inferred that during the process of UAV trajectory planning, the goal is to minimize the average AoI of users.
which represents the situation where the AoI value changes with the time slot. When $k=0$, the AoI is 0. As the time index increases, if the user's data remains uncollected, the AoI value will increase. However, if the user is within the transmission range of a UAV, the data is uploaded, and the AoI value is reset to zero. By adopting AoI as the metric, UAVs are encouraged to prioritize users with higher AoI values, and the goal is to minimize the mean AoI values of all users.

% Since there is no way to place a UAV so that all users are within the coverage of the UAV, the UAV needs to constantly move between users to gather information. 
% Since it is not possible to deploy enough UAVs to ensure all users are within the coverage area of UAVs, the UAVs need to continually move between users to gather information. Here we consider the UAV flying at a constant speed $v $ at a fixed altitude, but the UAV can optimize the flight trajectory by changing the angle of the flight direction with respect to the normal $\vartheta$, so that the update of the UAV's position from time slot $k$ to $k+1$ can be calculated as follows

Since it is not possible to deploy enough UAVs to ensure all users are within the transmission range area of UAVs, the UAVs need to continually move between users to collect data. The UAV flies at a constant speed $v$ and a fixed altitude. The flight trajectory of a UAV is controlled by changing the angle of the flight direction angles $\vartheta$ from the set $\langle 0^\circ, 45^\circ, 90^\circ, 135^\circ, 180^\circ, 225^\circ, 270^\circ, 315^\circ \rangle$. The UAV's position in time slot $k+1$ is updated as follows:
\begin{subequations}
\begin{align*}
    &x_{k+1,j}^{uav}=x_{k,j}^{uav}+v\cos\vartheta_{k,j},\\
    &y_{k+1,j}^{uav}=y_{k,j}^{uav}+v\sin\vartheta_{k,j},
\end{align*}
\end{subequations}
we consider distributed UAV flight trajectory optimization without a central controller. The UAVs have no prior knowledge about the users' positions, and need to detect the users' positions through their own sensors. Each UAV can also detect the positions of other UAVs in its vicinity.

% And since the UAV is distributed during the movement of the optimized flight trajectory, each UAV can also detect the positions of other UAVs in its vicinity. 

% Here we denote the detection range of the UAV by $r $, where $r \geq  d $.

% in this scenario, we consider distributed UAV flight trajectory optimization without a central controller. The UAVs, lacking prior knowledge about users' positions, employ radar sensors to detect user locations, thereby optimizing their flight trajectories to minimize the average AoI. As the UAVs operate distributively during the optimized flight path, each UAV can also detect the positions of other UAVs nearby through the same radar sensors. The detection range of each UAV is denoted by $r$, where $r \geq d$.

% Since in this paper we consider the distributed UAV flight trajectory optimization problem without a central controller, neither the central controller nor the satellite with a global observation field of view are considered. This means that the UAV has no prior knowledge about the user's position when flying the trajectory, and it needs to detect the user's position through its own sensors, so as to optimize the flight trajectory to minimize the average AoI. And since the UAV is distributed during the movement of the optimized flight trajectory, the UAV can only detect the position of other UAVs in its vicinity. Here we consider the detection range of the UAV as $r $, and since the sensing range of the sensor is usually larger than the communication range, in this paper $r \geq  d $.

\subsection{Problem Formulation}
% According to above description, the problem of UAVs flying trajectories optimization to minimize the average AoI of all users can be formulated as follows.
The problem of UAV trajectory optimization can be formulated as follows:
\begin{problem}\label{p1}
\begin{align}
&\min_{\bm{\vartheta}}\sum_{k=0}^{K}\sum_{i=1}^{N}a_{i}^{k},\label{obj}\\
    &s.t. \;\; x_{k+1,j}^{uav}=x_{k,j}^{uav}+v\cos\vartheta_{k}^{j},\tag{\ref{obj}a}\label{c1}\\
    &\qquad y_{k+1,j}^{uav}=y_{k,j}^{uav}+v\sin\vartheta_{k}^{j},\tag{\ref{obj}b}\label{c2}\\
    &\qquad a_{i}^{k} =
    \begin{cases}
        a_{i}^{k} + 1, & \text {$(x_{i}^{user},y_{i}^{user})\notin \mathcal{C}_{k}$},  \\
        0, & \text {$(x_{i}^{user},y_{i}^{user})\in \mathcal{C}_{k}$},
    \end{cases}\tag{\ref{obj}c}\label{c3}
\end{align}
\end{problem}
% \noindent where the above formula shows that within the entire time period $T$, it is necessary to minimize the average AoI for all users at every moment. By optimizing the trajectory of the UAVs, the aim is to minimize losses, thereby enhancing the data transmission efficiency of the entire model. Constraints \eqref{c1} and \eqref{c2} regulate the flying speed and the location of UAVs, while \eqref{c3} shows the update of the AoI. 
\noindent which shows that within the entire time period $K$, it is important to minimize the mean AoI values for all users in every time slot as much as possible. By optimizing the trajectory of the UAVs, the aim is to minimize losses, thereby enhancing the data transmission efficiency of the entire model. Constraints \eqref{c1} and \eqref{c2} regulate the flying speed and the location of UAVs, while \eqref{c3} shows the update of the AoI. 
\begin{lemma}\label{lemma-1}
    The Kolmogorov-Arnold representation theorem states that any multivariate function can be represented as a composition of univariate functions and summations
    \begin{align}
        f\left(x_1, \ldots, x_n\right)=\sum_{q=0}^{2 n} \Psi_q\left(\sum_{p=1}^n \Psi_{p, q}\left(x_p\right)\right).
    \end{align}
\end{lemma}
The number of \text{outer} functions $\Psi_q$ can be reduced to a single function $\Phi$ without a loss of representational complexity \cite{lorentz1962metric}. Furthermore, if the function space is constrained to the space of symmetric (permutation invariant) functions, then the set of \text{inner} functions $\Psi_{p, q}$ can be replaced with a single function $\phi$.

Due to the homogeneity of the UAVs, altering the UAV identifiers does not influence the output results. This characteristic leads to permutation invariance. Thus, according to Lemma \ref{lemma-1}, permutation invariant methods can be used for efficient computation.

\begin{figure*}[ht]
  \centering
  \includegraphics[width=1.7\columnwidth]{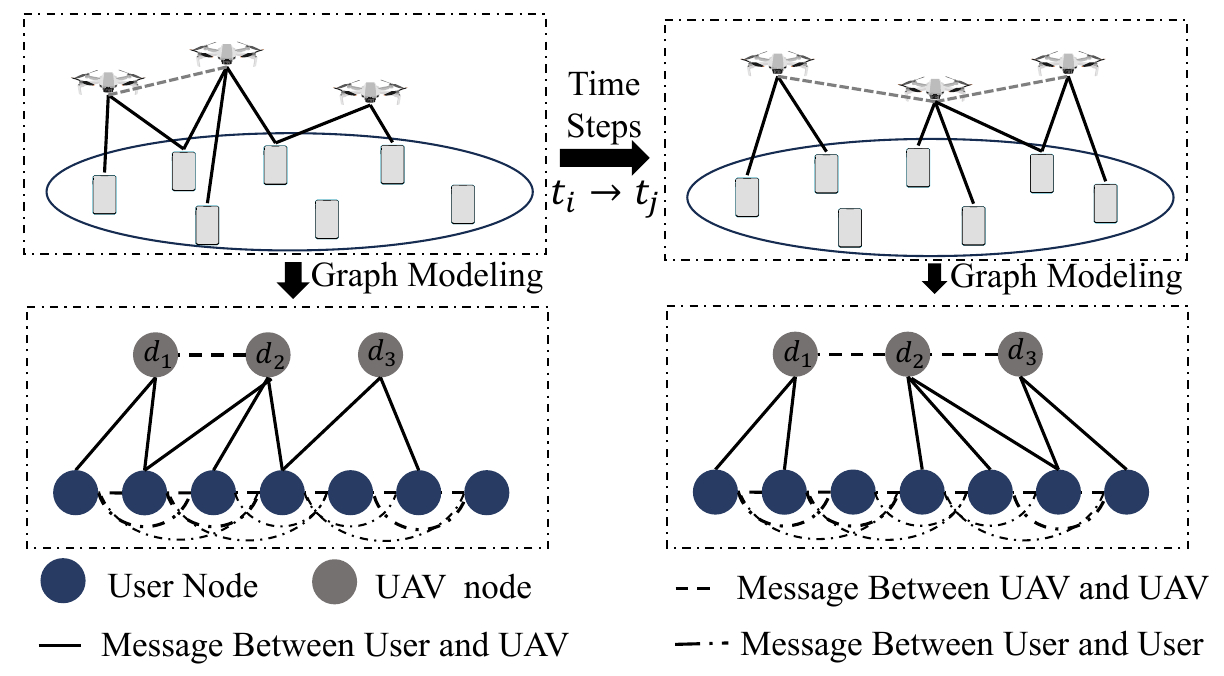}
   % \vspace{-9pt}
  \centering \caption{Illustration of graph models, with the movement of the UAVs, due to the change of the observation range and communication range, the graphical model corresponding to the wireless communication network composed of the UAVs and the users is also changing.} 
  \label{graph model}
   % \vspace{-9pt}
\end{figure*}

\section{QMIX-Based MARL Method}
\subsection{Dec-POMDP Modelling}
% Obviously, Problem \ref{p1} is a timing optimization problem. First, we need to analyze whether the problem has Markov property to determine whether reinforcement learning can be used to solve this problem. According to the relationship between the optimization variables in an optimization problem and the non-optimizable variables determined by the environment, we can define the state space and action space of the problem as follows.\\
% Problem \ref{p1} is a timing optimization problem. 

In Problem \ref{p1}, the environment state is not fully observable to the UAVs. UAVs can only observe the positions of users and UAVs within their detection range, and they can only partially observe the AoI for users within this range. This limited observation capacity necessitates a strategy that can operate under partial observability to optimize the UAVs' flight trajectories and resource allocation effectively. First, we need to analyze whether the problem has the Markov property to determine whether reinforcement learning can be used to solve it. We define the state space and action space of the problem as follows.\\
% $\bullet$\textbf{State}: The state space encapsulates all possible configurations of the environment, which includes the positions of all $N$ users, which are fixed, and the positions of all $M$ UAVs. Moreover, the AoI of all users are also included in the state space, since only when the UAV knows the user's AoI can it know how to optimize the flight trajectory to collect data from users with large AoI first, in order to reduce the mean AoI values.
$\bullet$\textbf{State}: The state space includes all possible configurations of the environment, such as the fixed positions of $N$ users, the positions of $M$ UAVs, and the AoI values of all users. 
\begin{align}
    S = \{(x_i, y_i)_{i=1}^{M}, (x_i, y_i, a_i)_{i=M+1}^{N+M}\},
\end{align}
where $(x_i, y_i)$ represents the positions of all $M$ UAVs with $1 \leq i \leq M$, and $(x_i, y_i, a_i)$ represents the positions and AoI values of all $N$ users with $M+1 \leq i \leq M+N$.\\
% $\bullet$\textbf{Action}: Since the optimization variable in Problem \ref{p1} is only $\vartheta$, the action space is the flight direction of all UAVs that is $\{\vartheta^1,\cdots,\vartheta^M\}$.\\
$\bullet$\textbf{Action}: Since the optimization variable in Problem \ref{p1} is ${\bm{\vartheta}}$, the action space is the flight direction of all UAVs, i.e., $\{\vartheta^1,\cdots,\vartheta^M\}$.
The position of the UAV in the next time slot depends only on the position of the UAV and flight direction in the current time slot. According to equation \eqref{c3}, the user's AoI in the next time slot also depends only on the user's AoI values at the current time and whether the user will be within the transmission range of the UAVs in the next time slot. 

The values of all variables in the state space in the next time slot depend only on the values of variables in the current time slot and the current time slot's action. Thus, Problem \ref{p1} can be modeled as a Markov Decision Process (MDP). However, due to the limited detection range of each agent, that is, the UAV, the actual detection range of each agent does not match the size of the state space. The observation of each agent is defined as follows:\\
$\bullet$\textbf{Observation}: The observation $o_i$ of the agent $i$ includes the locations of the other UAVs and users within a detection range of $r$, as well as the user's AoI. The $o_i$ can be expressed as follows.
\begin{align*}
    o_{i} = &\left\{(x_{j}^{user},y_{j}^{user},a_{k,j})|(x_{j}^{user},y_{j}^{user})\in \mathcal{D}_{k,i} \right\}\\
    &\quad\bigcup\left\{(x_{k,j}^{uav},y_{k,j}^{uav})|(x_{k,j}^{uav},y_{k,j}^{uav})\in \mathcal{D}_{k,i} \right\}.
\end{align*}
As mentioned above, each UAV cannot fully observe the state information of the environment. Therefore, Problem \ref{p1} can be modeled as a Dec-POMDP.

% The objective of Problem \ref{p1} is to minimize the mean AoI values, so the reward can be defined as follows.\\
% $\bullet$\textbf{Reward}: The instantaneous reward for taking actions is defined as the negative mean AoI values of all users, i.e., the negative of the objective function value in equation \eqref{obj}. By maximizing the reward, mean AoI values can be minimized. Moreover, since the reward is defined as the instantaneous reward for taking a specific action at the current moment, the reward should be the opposite of the instantaneous AoI, thus
% \begin{align}
%     r_{t}=-\sum_{i=1}^{N}a_{i}^{t}.
%     \label{instant-reward}
% \end{align}

% The objective of Problem \ref{p1} is to minimize the mean AoI values, so the reward can be defined as follows.\\
% $\bullet$\textbf{Reward}: The instantaneous reward for taking actions is defined as the negative mean AoI values of all users, i.e., the negative of the objective function value in equation \eqref{obj}. By maximizing the reward, mean AoI values can be minimized. Moreover, since the reward is defined as the instantaneous reward for taking a specific action at the current moment, the reward should be the opposite of the instantaneous AoI, thus
% \begin{align}
%     r_{t}=-\sum_{i=1}^{N}a_{i}^{t}.
%     \label{instant-reward}
% \end{align}

The objective of Problem \ref{p1} is to minimize the mean AoI values, so the reward can be defined as follows.\\
$\bullet$\textbf{Reward}: RL is a machine learning approach that learns decision-making strategies by interacting with the environment to maximize expected rewards. In our problem, the objective is to minimize the mean AoI values of users. To achieve this, we use the negative of users' AoI values as rewards and maximize this reward by controlling agents to minimize the AoI values. The instantaneous reward for taking actions is defined as the negative mean AoI values of all users, i.e., the negative of the objective function value in equation \eqref{obj}. By maximizing this negative reward, we are able to minimize the mean AoI values. 
% The instantaneous reward $r_{t}$ can be defined as follows.
\begin{align}
    r_{k}=-\sum_{i=1}^{N}a_{i}^{k}.
    \label{instant-reward}
\end{align}

Equation \eqref{instant-reward} is denoted as the instantaneous reward generated after the agents take joint actions in time slot $k$.

Due to the fact that most MARL algorithms use centralized training and decentralized execution (CTDE), we can represent the performance of the entire system during training using the joint action-value function, $Q_{tot}^{k}$. This function is estimated by combining the local Q-values of each agent and reflects the future joint action-value in time slot $k$. $Q_{tot}^{k}$ includes the discounted sum of all future rewards, guiding agents to optimize both current performance and future impacts, thereby achieving long-term goals. According to the Bellman equation \cite{goodfellow2016deep}, in time slot $k$, maximizing the joint action-value function $Q_{tot}^{k}$ is achieved by taking actions ${\bm{\vartheta}_{k}} = \{\vartheta^{i}_{k}, \cdots, \vartheta^{M}_{k}\}$, which are selected based on the local Q-values of each agent. This problem can be modeled as follows.
\begin{problem}
\begin{align}
    &Q_{tot}^{k} = -\sum_{i=1}^{N} a_{i}^{k}+\gamma Q_{tot}^{k+1},\label{q-tot}\\
    &s.t. \;\; x_{k+1,j}^{uav}=x_{k,j}^{uav}+v\cos\vartheta_{k}^{j}, \;j\in\{1,\cdots,M\},\tag{\ref{q-tot}a}\label{c-q-1}\\
    &\qquad y_{k+1,j}^{uav}=y_{k,j}^{uav}+v\sin\vartheta_{k}^{j},\; j\in\{1,\cdots,M\},\tag{\ref{q-tot}b}\label{c-q-2}\\
    &\qquad a_{i} =
    \begin{cases}
        a_{i} + 1, & \text {$(x_{k,i}^{user},y_{k,i}^{user})\notin \mathcal{C}_{k}$},  \\
        0, & \text {$(x_{k,i}^{user},y_{k,i}^{user})\in \mathcal{C}_{k}$},
    \end{cases} \tag{\ref{q-tot}c}\label{c-q-3}\
\end{align}
\end{problem}
\noindent where $\gamma$ is the discount factor, $Q_{tot}^{k}$ estimates the future joint action-value in time slot $k$, $Q_{tot}^{k+1}$ estimates the future joint action-value in time slot $k+1$.

\subsection{QMIX Method}
% In MARL, algorithms are mainly divided into two categories: value decomposition and policy gradient. Here, we focus on value decomposition-based MARL algorithms. These algorithms primarily involve two components: the joint action-value function and the local action-value function. The joint action-value function (global Q-value) is used to evaluate the overall performance of the entire agent group. Each agent's local action-value function (local Q-value) estimates the value of each possible action based on the agent's own state, and then selects the optimal action through $\epsilon$-greedy.
% In MARL, algorithms can be broadly categorized in various ways, including value decomposition methods and policy gradient methods. Here, we focus on value decomposition-based MARL algorithms. These algorithms often involve two key components: the joint action-value function (global Q-value) and the local action-value functions (local Q-values). The joint action-value function is used to evaluate the overall performance of the entire agent group. Each agent has its own local action-value function, which estimates the value of each possible action based on both its own state and potentially the states or actions of other agents. The local action-values are then combined in a specific manner to form the global Q-value. Agents typically select their actions using a strategy such as $\epsilon$-greedy, balancing exploration and exploitation.
In MARL, algorithms are mainly divided into two categories: value decomposition and policy gradient. Here, we focus on value decomposition-based MARL algorithms. These algorithms often involve two key components: the joint action-value function (global Q-value) and the local Q-value functions. The global Q-value is used to evaluate the overall performance of the entire agent group. Each agent has its own local Q-value function, which estimates the value of each possible action based on both its own state and potentially the states or actions of other agents. The agents then select the optimal action using the \(\epsilon\)-greedy strategy \cite{sutton2018reinforcement}.
% In MARL, algorithms are mainly divided into two categories: value decomposition and policy gradient. Here, we focus on value decomposition-based MARL algorithms. These algorithms often involve two key components: the joint action-value function (global Q-value) and the local action-value functions (local Q-values). The global Q-value is used to evaluate the overall performance of the entire agent group. Each agent has its own local Q-value, which estimates the value of each possible action based on both its own state and potentially the states or actions of other agents. The agents then select the optimal action using the \(\epsilon\)-greedy strategy \cite{sutton2018reinforcement}.
% \begin{align}
%     a_t =
%     \begin{cases}
%     \underset{a}{\arg\max} \, Q(s_t, a) & \text{with probability } 1 - \epsilon, \\
%     \text{random action from } A & \text{with probability } \epsilon,
%     \end{cases} \label{greedy}
% \end{align}
% where $\epsilon$ determines the probability of exploring a random action, $Q(s_t, a)$ is the action-value function for state $s_t$ and action $a$, and $A$ is the set of all possible actions.

Early MARL algorithms, such as Value Decomposition Networks (VDN) \cite{sunehag2017value}, are characterized by the assumption that the sum of local Q-values from all agents equals the global Q-value. This assumption suggests that each agent can enhance the global Q-value by maximizing its local Q-value. However, this oversimplified method overlooks the fact that individual agents can have varying impacts on overall performance due to their distinct local characteristics. Furthermore, the direct aggregation in VDN poses challenges for agents to effectively learn how to collaborate with one another. In contrast, the QMIX algorithm significantly alleviates the limitations of VDN in aggregating local Q-values by using a more flexible mixer network. By introducing a mixer network, QMIX nonlinearly combines the local Q-values of agents, enabling them to understand how their actions influence the global Q-value more effectively. In the training phase, QMIX uses the mixer network to non-linearly combine the local Q-values of different agents. Through the mixer network, each agent learns its local Q-value and understands its impact on the global Q-value. This process enables each agent to learn how to cooperate with other agents to enhance overall performance. In the inference phase, after learning the impact of its local Q-value on the global Q-value, each agent can directly use its local Q-value to select actions without relying on the mixer network again, thereby enabling distributed inference.

The QMIX algorithm ensures that when performing an $\text{argmax}$ operation on the global Q-value $Q_{tot}$, the resulting joint action set $\bm{\vartheta}$ remains consistent with the combination of actions obtained by performing $\text{argmax}$ on each local Q-value $Q_{i}$. In other words, the local optimal actions chosen by each agent are precisely a subset of the global optimal actions. This property can be expressed as follows:
\begin{align}
    \underset{\bm{\vartheta}}{\arg\max}\quad Q_{tot} = \left(\begin{array}{cc}
\underset{\vartheta_1}{\arg \max } & Q_{1} \\
\cdots\\
\underset{\vartheta_M}{\arg \max } & Q_M
\end{array}\right).
\label{qmix2}
\end{align}

The operation defined by equation \eqref{qmix2} can be extended to a broader space of monotonic functions. By ensuring monotonicity through partial derivatives, i.e., when the local Q-value of each individual agent increases, the global Q-value also increases or remains constant, thereby achieving maximization of the global Q-value. This condition is specifically formulated as:
\begin{align}
\frac{\partial Q_{tot}}{\partial Q_{i}} \geq 0, \quad \forall i \in {1, \ldots, M}. \label{qmix}
\end{align}

Through this method, we can guarantee the attainment of desired outcomes in joint action selection, where the increase in Q-value for each agent contributes to optimizing the overall system performance. The challenge is how to ensure the validity of equation \eqref{qmix}. Fortunately, a multi-layer perceptron (MLP) can be viewed as a superposition of $L$ non-linear layers, the $i-$th layer can is computed as follows:
\begin{align}
    x^{i} = \sigma^{i}(W^{i}x^{i-1}+b^{i}), \label{non-linear}
\end{align}
where $\sigma^{i}(\cdot)$ is a non-linear activation function used in the $i-$th layer, and $W^{i}$ and $b^{i}$ are trainable parameters in the $i-$th layer. To ensure that equation \eqref{non-linear} is monotonically increasing with respect to $x$, two conditions need to be satisfied: 1) $W^{i}$ needs to be larger than 0, and 2) $\sigma^{i}(\cdot)$ needs to be monotonically increasing. The $W^{i}$ larger than 0 can be achieved by employing the absolute function $\psi(\cdot)$, and ensuring the activation function's monotonically increasing property can be guaranteed by setting $\sigma^{i}(\cdot)$ as the ReLU function. 

The challenge is to determine the parameters of $\bm{W}$ and $\bm{b}$. The purpose of $\bm{W}$ and $\bm{b}$ is to compute the global Q-value at the current step by extracting the Q-value from the local Q-Network at the same step. Thus, $\bm{W}$ and $\bm{b}$ should be related to the current state or observation, rather than being fixed values independent of the state. Naively training a monotonically increasing neural network (NN) for each agent to capture the relationship between its local Q-value and the global Q-value can satisfy the requirements of Equation (10). However, this method increases the storage overhead and limits the amount of data available to train each NN individually, thereby affecting the overall training efficacy. Fortunately, Lemma 1 states that for a permutation-invariant equation, the same $\phi(\cdot)$ can be used to extract local features. Problem 1 is permutation-invariant, so instead of training an NN for each agent, we can train a parameter-sharing NN, $\Psi_{\text{inner}}$, as $\phi(\cdot)$ in Lemma 1. This NN comprises two cascaded parts: the first part takes the local Q-value of the $i$-th agent and the current state as inputs and outputs parameters $W_{\text{inner}}$ and $b_{\text{inner}}$ used to extract the local Q-value feature. These parameters, $\psi(W_{\text{inner}})$ and $b_{\text{inner}}$, serve as inputs to another monotonically increasing NN $\phi(\cdot)$ with ReLU activation, which outputs a feature map of the local Q-value relative to the global Q-value. After aggregating all feature maps according to Lemma 1, they are fed into another NN, $\Psi_{\text{outer}}$, which has a structure similar to $\Psi_{\text{inner}}$. $\Psi_{\text{outer}}$ uses the aggregated feature map and the current state to output $W_{\text{outer}}$ and $b_{\text{outer}}$. These parameters, $\psi(W_{\text{outer}})$ and $b_{\text{outer}}$, are then used in another monotonically increasing NN $\Phi(\cdot)$ with ReLU activation to estimate the global Q-value. The entire computational procedure can be formulated as follows:
\begin{align}
    &Q_{tot} = \Phi\left(\bigoplus_{i=1}^{M} \phi\left(Q_{i},s;\bm{W}^{i}_{\text{inner}},\bm{b}^{i}_{\text{inner}}\right); \bm{W}_{\text{outer}},\bm{b}_{\text{outer}}\right), \label{eq:mixer-1} \\
    &[\bm{W}_{\text{outer}},\bm{b}_{\text{outer}}] = \Psi_{\text{outer}}\left(\bigoplus_{i=1}^{M} \phi\left(Q_{i},s;\bm{W}^{i}_{\text{inner}},\bm{b}^{i}_{\text{inner}}\right),s\right),\label{eq:mixer-2}\\
    &[\bm{W}^{i}_{\text{inner}},\bm{b}^{i}_{\text{inner}}] = \Psi_{\text{inner}}\left(Q_i,s\right),\label{eq:mixer-3}
\end{align}
where $\bigoplus(\cdots)$ is the permutation invariant operator, i.e., summation or multiplication. It should be emphasized that although we use the global state $s$ in mixer NN in equations \eqref{eq:mixer-1}-\eqref{eq:mixer-3}, this is only used during training to improve the training accuracy. In actual operation, each agent only needs to use the local Q-Network estimate at the local Q-Network to select the action with the largest local Q-value to execute. Because through the training of NN and equation \eqref{qmix}, it is guaranteed to select the action with the largest local Q-value to maximize the global Q-value. This way, our algorithm can achieve CTDE.

% \begin{figure*}[ht]
%   \centering
%   \includegraphics[width=1.9\columnwidth]{fig/第一张系统图2.pdf}
%    % \vspace{-9pt}
%   \centering \caption{Agent Communication Structure Model in the Qedgix Algorithm.} 
%   \label{system model}
%    % \vspace{-9pt}
% \end{figure*}

In the training procedure, all local Q-networks $\mathcal{Q}_{i},\forall i \in\{1,\cdots,M\}$ and the mixer network $\phi$ are regarded as a unified network $\mathcal{Q}$, whose input is the observation of all agent and the state, and output is the estimated global Q-value $Q_{tot}$. During the training process, in order to ensure algorithm convergence and enhance algorithm performance stability, similar to \cite{goodfellow2016deep}, when training the QMIX network, the target network $\mathcal{Q}_{target}$ is used. The parameters of this target network $\mathcal{Q}_{target}$ remain static during backpropagation, serving as a stable reference. Periodically, the parameters of the network $\mathcal{Q}$ are copied into $\mathcal{Q}_{target}$, aligning the target network with the latest learned information without directly participating in the training process. The training loss of the unified network $\mathcal{Q}$ is as follows:
% \begin{align}
%     &\bm{L\left(\vartheta \right)} = \left(y_{tot}-\mathcal{Q}\left(\bm{\vartheta}_{t},\bm{o}_{t};\bm{s}_{t}\right)\right)^2,\\
%     &y_{tot}=r+\gamma \max_{\bm{\vartheta}_{t+1}} \mathcal{Q}_{target}\left(\bm{\vartheta}_{t+1},\bm{o}_{t+1};\bm{s}_{t+1}\right)
% \end{align}
\begin{align}
    &\bm{L\left(\vartheta \right)} = \left(y_{tot}-\mathcal{Q}\left(\bm{\vartheta}_{k},\bm{o}_{k};\bm{s}_{k}\right)\right)^2,\\
    &y_{tot}=r+\gamma \max_{\bm{\vartheta}_{k+1}} \mathcal{Q}_{target}\left(\bm{\vartheta}_{k+1},\bm{o}_{k+1};\bm{s}_{k+1}\right).
\end{align}
According to the above equations, the parameters $\bm{\theta}_{i}$ for local Q-network $\mathcal{Q}_i$ can be updated as follows:
\begin{align}
    \bm{\theta}_{i} &= \bm{\theta}_{i} - lr\frac{\partial \bm{L(\vartheta)}}{\partial \bm{\theta}_{i}},\notag\\
    &= \bm{\theta}_{i}  - lr\frac{\partial \bm{L(\vartheta)}}{\partial \mathcal{Q}\left(\bm{\vartheta}_{k},\bm{o}_{k};\bm{s}_{k}\right)}\frac{\partial \mathcal{Q}\left(\bm{\vartheta}_{k},\bm{o}_{k};\bm{s}_{k}\right)}{\partial \mathcal{Q}_{i}(\vartheta_{i},o_{i})},\notag\\
    &=\bm{\theta}_{i} - 2lr(y_{tot}-\mathcal{Q}\left(\bm{\vartheta}_{k},\bm{o}_{k};\bm{s}_{k}\right))\frac{\partial \sigma(\psi(W)Q_{i}+b)}{\partial (\psi(W)Q_{i}+b)}\psi(W),
\end{align}
where $lr$ is the learning rate. From the above equation for updating $\bm{\theta}_{i}$, the global state information is only utilized when computing $Q_{tot}$. This method provides a more accurate gradient direction for $\frac{\partial \bm{L}(\vartheta)}{\partial \bm{\theta}_{i}}$ by directly calculating the gradient of the global reward. This calculation is performed with respect to the parameters of each local Q-network, rather than updating the local Q-network parameters using only the local reward. The agent does not use the global state information in the process of evaluating the impact of local actions on the global reward. Therefore, it can implement distributed decisions using only the information it observes.

\section{Graph Structure Introduction And Optimization}
In this section, we will introduce how to use GNN to enhance the performance of QMIX.
\subsection{Explanation of EdgeConv in Communication Structures}
The QMIX algorithm, known for its exceptional performance, is classic in the field of MARL. However, due to the limited capabilities of the UAVs' detection sensors, they are unable to acquire complete global state information. When the QMIX algorithm is deployed for training on this task, performance will be subject to certain limitations. The neural networks of agents in the QMIX algorithm adopt a Gated Recurrent Unit (GRU) structure. This simplified structure restricts the algorithm's ability to learn the collaborative relationships among UAVs and the interactions between UAVs and users within the environment.

\begin{figure*}[ht]
  \centering
  \includegraphics[width=1.90\columnwidth]{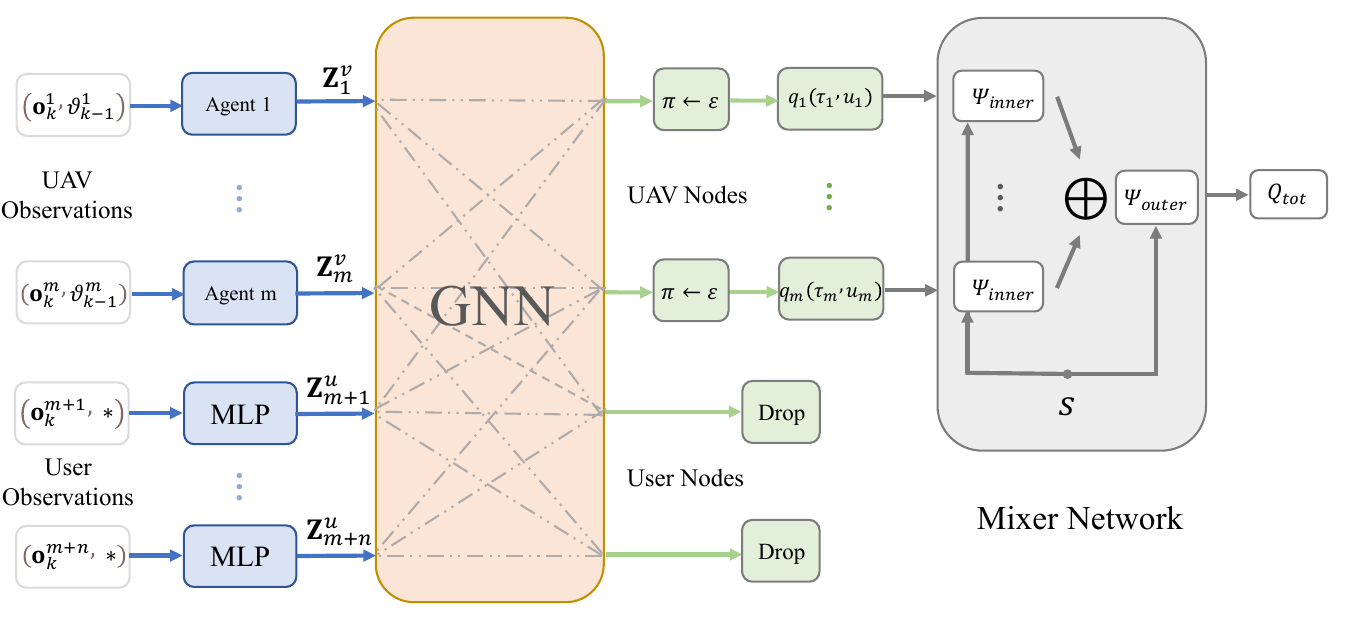}
   % \vspace{-9pt}
  % \centering \caption{Introduction to the Qedgix Algorithm Framework. The training process of the Qedgix algorithm involves several steps: UAV nodes input observed information and actions into corresponding agent networks to generate node features for the GNN. User nodes input observed information into specialized MLP feature extraction networks, also producing node features suitable for the GNN. Subsequently, these features undergo distributed processing through the GNN to obtain updated node representations. Since user nodes do not need to move, they discard the corresponding processing results as they do not require $\epsilon$-greedy strategy to obtain appropriate action instructions. In contrast, information from UAV nodes generates Q-values through the $\epsilon$-greedy strategy, which then enter the Mixer Network for aggregation, producing global Q-values to optimize the entire system. The Mixer Network is used for GNN training and discarded during inference to enable distributed execution solely through the GNN.} 
  \centering \caption{Introduction to the Qedgix Algorithm Framework: The framework utilizes a GNN to extract features from UAVs and users, evaluating how various UAV flight directions affect the average AoI through the outputs of corresponding UAV nodes. The Mixer Network is used during training but discarded during inference.} 
  \label{system model}
   % \vspace{-9pt}
\end{figure*}

To address the challenges posed by the simplistic network structure of the QMIX algorithm in multi-agent task environments, which hampers the learning of complex inter-agent information, we propose an improved version of the QMIX algorithm—named Qedgix. This new algorithm integrates the EdgeConv graph neural network \cite{wang2019dynamic}, leveraging its capacity for efficient feature extraction and representational learning in graph-structured data, thereby augmenting the perceptual abilities of UAV agents. In particular, this improvement enables each agent to accurately comprehend the complex relationships with other agents, thereby enhancing the accuracy of decision-making and facilitating more coordinated collaboration, which in turn boosts the efficiency of task completion in multi-agent environments. The reason we can use GNN in the QMIX algorithm is that each agent selects appropriate actions based on its local Q-network, and subsequently computes the global Q-value \(Q_{tot}\) through a mixer network, achieving holistic optimization of multi-agent systems. This process can be delineated into two tasks: node regression and graph regression. Specifically, each agent's action selection via the local Q-network constitutes the node regression task. The information output by the GNN nodes is a vector of Q-values corresponding to the actions each agent can take. These vectors represent the impact of different flight directions on optimizing the AoI. While the computation of $Q_{tot}$ represents the graph regression task. However, given the distinct optimization objectives of these tasks during training, it is challenging to simultaneously address both with GNN alone. Therefore, we propose the GNN-QMIX method, which integrates GNN to enhance inter-agent information exchange, while retaining the mixer network from QMIX algorithm.

Fig.~\ref{system model} illustrates the agent structure design in the Qedgix algorithm, which comprises several key components: an MLP, a GRU for processing the observation history \cite{chung2014empirical}, and the EdgeConv graph neural network to enhance information exchange. The training process involves several steps: UAV nodes input observed information and actions into corresponding agent networks to generate input  node features for the GNN. User nodes input observed information into specialized MLP networks, also producing input node features for the GNN. These features undergo distributed processing through the GNN to update node representations. User nodes drop the results as they do not need to make actions, while UAV nodes generate local Q-values using the $\epsilon$-greedy strategy, which then enter the Mixer Network for aggregation, producing global Q-values $Q_{tot}$ to optimize the system. The Mixer Network is used for GNN training and discarded during inference, enabling distributed execution solely through the GNN. The design of this algorithmic structure optimizes the information exchange between agents and the overall decision-making process, thereby improving the efficiency and quality of decisions made by the algorithm. In the GNN model involving UAVs and users, they are treated as nodes in the graph neural network, with wireless channels between them modeled as edges. The exchange of information between UAVs and users enhances global resource and trajectory optimization, thus forming a bidirectional graph.

The messaging mechanism of EdgeConv is as follows:
\begin{align}
    &X_{i}^{'} ={\sum_{j\in \mathcal{N}(i)}f\left(X_{i}\parallel X_{j}-X_{i}  \right)},\quad\quad\quad\;   \mathbf{A}_{i,j}\neq0 \label{message:1}\\
    &X_{m} = W_1[o_{m}, h_{m}],\quad\qquad\qquad\quad\qquad\,\,\,\forall m \in \mathcal{V}^{v} \label{message:2}\\
    &X_{n}=W_2[o_{n}],\qquad\qquad\qquad\qquad\qquad\,\,\,\,\forall n \in \mathcal{V}^{u} \label{message:3}
\end{align}
where $\mathcal{V}^{v}$ and $\mathcal{V}^{u}$ are the sets of UAVs and users respectively. $\mathcal{N}(i)$ represents the other nodes that are within the detection range of node $i$. EdgeConv updates node attributes by means of a symmetric operation $\left(X_{i} \| X_{j}-X_{i}  \right)$, which integrates both the features of the node itself and those of its adjacent nodes. This process employs a nonlinear mapping function $f(\cdot)$, equipped with a series of learnable parameters, to transform and aggregate the obtained features. $W_1$ is used to extract information from UAV node features $o$ and to store hidden features $h$ generated from past observations, while $W_2$ is used to extract information from user node features and to normalize the dimensions of node feature inputs.
% $W_1$ and $W_2$ are different training parameter matrices. $W_1$ is used for feature extraction of UAV observations and preservation of historical observations, and $W_2$ is used for feature extraction of users observations and unification of node feature input dimensions.
% W1 and W2 are different training parameter matrices. W1 is used for feature extraction of drone observations and storage of historical observation values, while W2 is used for feature extraction of user observations and normalization of node feature input dimensions.

\subsection{Graph Formulation}
Training and interacting with the environment are crucial for reinforcement learning algorithms. Through the interaction, the policy network can evaluate and adjust the behavior of the agent based on feedback from the reward function, gradually improving the agent's performance on the task. In the interaction process, the key information required for algorithm training includes not only the magnitude of the reward values but also the state information observed by the agent. The following sections will provide a detailed explanation of how the agent network accurately processes state information and ultimately makes the correct actions.

The node features and edge attributes of the graph neural network are constructed in real time in the environment. The edge attributes is constructed by the adjacency matrix. Already set in the system model, the size of the adjacency matrix is $\left(M + N\right)\times\left(M + N\right)$. The elements in the adjacency matrix $\mathbf{A}\in \mathbb{R}^{(M+N)\times (M+N)}$ are either 0 or 1, determined by whether the distance between two nodes is within the detection range of the UAV. If the distance between two nodes falls within the detection range, the nodes are considered to be related. In this scenario, the corresponding values in the adjacency matrix, $\mathbf{A}_{(i,j)}$ and $\mathbf{A}_{(j,i)}$, are set to 1. If the distance between the nodes exceeds the detection range, these values are set to 0.
\begin{align}
    \mathbf{A}_{(i,j)} =\begin{cases}
    1, & \left({x}_{i},{y}_{i}\right) \in \mathcal{D}_{j},\\
    0, & \left({x}_{i},{y}_{i}\right)\notin \mathcal{D}_{j},
    \end{cases}
\end{align}
where the indices $i, j \in \{1, \ldots, M+N\}$. The indices from $1$ to $M$ represent the UAV nodes, where as the indices from $M+1$ to $M+N$ represent the user nodes.

As previously discussed, the construction of the adjacency matrix in the GNN has been addressed. In the following, we elaborate on the formation of the input node features for the GNN. The input node features of the GNN are derived from the observed information. This observed information can be expressed as a matrix, where each row corresponds to a node and each column corresponds to a feature. We represent the observed information as $\mathbf{O} = \{\mathbf{O}^{v}, \mathbf{O}^{u}\}$.

The matrix $\mathbf{O}^{v}$ represents the UAVs' information (relative coordinates) observed by each node, and its dimensions are $\mathbf{O}^{v} \in \mathbb{R}^{\left(M+N\right)\times M \times 2}$. The UAVs' relative coordinates observed by the $i$-th node can be expressed as:
\begin{align}
    \mathbf{O}^{v}_{(i,:,:)} &= \left[\left[\Delta x_{i,1}, \Delta y_{i,1}\right], \cdots, \left[\Delta x_{i,M}, \Delta y_{i,M}\right]\right].
\end{align}

Similarly, the matrix $\mathbf{O}^{u}$ represents the users' information (relative coordinates and user AoI) observed by each node, and its dimensions are $\mathbf{O}^{u} \in \mathbb{R}^{\left(M+N\right)\times N \times 3}$. The users' information observed by the $i$-th node can be expressed as:
\begin{align}
    \mathbf{O}^{u}_{(i,:,:)} &= \left[\left[\Delta x_{i,1}, \Delta y_{i,1}, a_{i,1}\right], \cdots, \left[\Delta x_{i,N}, \Delta y_{i,N}, a_{i,N}\right]\right].
\end{align}

If a node $i$ is unable to observe an entity (such as a UAV or user) due to limited detection range, the observation value for that entity is set to zero.
% The GNN graph node features $\mathbf{Z}=\{\mathbf{Z}^{v},\mathbf{Z}^{u}\}$ are composed of user and UAV node features. The UAV node features are obtained by combining real-time observed information from the environment with hidden features generated in the previous time slot through a GRU recurrent neural network, while user node features are collected directly from the environment. To ensure consistent feature dimensions across input nodes for EdgeConv, an MLP fully connected layer is used.
\begin{align}
\mathbf{Z}_{i}^{v} &= \text{GRU}(\mathbf{O}_{(i,:,:)}, h^{v}_{k-1}), \label{nodev}\\
\mathbf{Z}_{i}^{u} &= \text{MLP}(\mathbf{O}_{(i,:,:)}).\label{nodeu}
\end{align}

The GNN graph input node features $\mathbf{Z}=\{\mathbf{Z}^{v},\mathbf{Z}^{u}\}$ are composed of user and UAV node features. As seen in equation \eqref{nodev}, the UAV node features are obtained by combining real-time observed information from the environment with hidden features generated in the previous time slot through a GRU recurrent neural network. As demonstrated in equation \eqref{nodeu}, the user node features are collected directly from the environment. To ensure consistent feature dimensions across input nodes for EdgeConv, an MLP fully connected layer is used.

\begin{algorithm}[h]
    \caption{Inference Procedure of Qedgix Algorithm}
    \label{alg:inference_enhanced_qmix}
    \begin{algorithmic}[1]
    \renewcommand{\algorithmicrequire}{ \textbf{Input:}}
    \renewcommand{\algorithmicensure}{ \textbf{Output:}}
    \REQUIRE UAV observations $\{\mathbf{O}^{1}, \mathbf{O}^{2}, \ldots, \mathbf{O}^{m} \}$, user observations $\{\mathbf{O}^{m+1}, \mathbf{O}^{m+2}, \ldots, \mathbf{O}^{m+n} \}$, adjacency matrix $\mathbf{A}$
    \ENSURE Actions for UAVs $\bm{\vartheta} = \{\vartheta^1, \vartheta^2, \ldots, \vartheta^m\}$
    
    \STATE Initialize environment and set initial observations $\{\mathbf{O}^{1}, \mathbf{O}^{2}, \ldots, \mathbf{O}^{m+n}\}$

    % \FOR{\added{each time step $k$}}
    \FOR{each time step $k$}
        \FOR{each UAV $i = 1$ to $m$}
            \STATE Input UAV observation $\mathbf{O}^{i}_{k}$ and previous hidden state $h^{i}_{k-1}$ into GRU layer
            \STATE Process through GRU layer to obtain hidden state $h^{i}_k$
        \ENDFOR
        
        \FOR{each user $j = m+1$ to $m+n$}
            \STATE Input user observation $\mathbf{O}^{j}_{k}$ into MLP layer
            \STATE Process through MLP layer to obtain hidden state $h^{j}_k$
        \ENDFOR
        
        \STATE Construct adjacency matrix $\mathbf{A}$ based on detection range
        
        \FOR{each node $i$ in the graph}
            \STATE Aggregate information from neighboring nodes using EdgeConv
            \STATE Update node attributes $X'_i = \sum_{j \in \mathcal{N}(i)} f(X_i \parallel X_j - X_i)$ 
        \ENDFOR
        
        \FOR{each UAV $i = 1$ to $m$}
            \STATE Calculate Q-value $Q_i$ from the updated node attributes
            \STATE Determine action $\vartheta^i = \arg\max Q_i$
        \ENDFOR
        
        \STATE Execute actions $\bm{\vartheta}$ and update environment
    \ENDFOR
    
    \end{algorithmic}
\end{algorithm}

\subsection{Feasibility of the methodology}
% The EdgeConv of GNN method plays a significant role in the Qedgix algorithm, primarily serving to refine the feature representation of each node by aggregating information from adjacent nodes. Specifically, the new feature of any node $v_i \in V$ obtained after the EdgeConv operation is denoted as $x'$. As indicated by Equation \eqref{eq:propagate}, this aggregation operation is accomplished through summation, implying that the central node does not need to consider the order of input from its neighbor nodes during the information processing phase. The detailed illustration is as follows:
The EdgeConv operation plays a significant role in the Qedgix algorithm, primarily serving to refine the feature representation of each node by aggregating information from adjacent nodes. Specifically, the new representation of any node $v_i \in V$ obtained after the EdgeConv operation is denoted as $x'$. As indicated by equation \eqref{message:1}, this aggregation operation is accomplished through summation, implying that each node does not need to consider the order of input from its adjacent nodes during the feature extraction phase. The detailed illustration is as follows:
\begin{enumerate}
    % \item \textit{Permutation invariance of the aggregation operation}:
    
    % The aggregation operation sums over all neighbor feature differences. Since the summation operation is symmetric, for any permutation $\pi$, we have:
    % \begin{equation}
    % \sum_{j \in \mathcal{N}(i)}(.) = \sum_{\pi(j) \in \mathcal{N}(i)}(.)
    % \end{equation}

    % Therefore, the aggregation operation is not affected by the order of neighbor nodes, thereby permutation invariance.
    \item \textit{Permutation invariance of the aggregation operation}:
    
    The aggregation operation computes the sum of the differences between the features of the node and those of its adjacent nodes, contributing to the update of the node's feature representation. Since the summation operation is symmetric, the aggregation operation is not affected by the order of adjacent nodes, thereby maintaining permutation invariance.
    
    \item \textit{Permutation invariance of the nonlinear function}:
    
    For each node $v_i$ and its adjacent node $v_j$, the difference of feature is denoted as $X_j - X_i$. This calculation of disparity is inherently permutation invariant, as it focuses solely on the features between nodes, independent of the sequence of adjacent nodes. For any permutation $\pi$, we have:
    \begin{equation}
    h(\pi(X_i), \pi(X_j) - \pi(X_i)) = h(X_i, X_j - X_i).
    \end{equation}
\end{enumerate}
This ensures that the computation result of $x'$ remains invariant regardless of the input sequence of adjacent nodes, thereby achieving permutation invariance.

% Furthermore, as indicated by Equation \eqref{eq:mixer}, the mixer network of QMIX also exhibits permutation invariance while aggregating local Q values from various agents. This demonstrates the feasibility and rationality of integrating graph neural network structures within the QMIX framework in the Qedgix algorithm. Namely, the graph neural network relied upon by drones for collecting information from adjacent nodes (other drones or users) does not depend on the specific order of nodes, enabling effective information transmission and extraction. This approach underscores the powerful capability of graph neural networks in processing dynamic and complex network structures, especially in flexibly adapting to collaborative tasks and decision-making processes within multi-UAV systems without the explicit need to consider node ordering, thus ensuring the model's adaptability to dynamic environmental changes.

Furthermore, as indicated by equation \eqref{eq:mixer-1}, the mixer network of QMIX also exhibits permutation invariance while aggregating local Q values from agents. This demonstrates the feasibility of integrating graph neural network within the QMIX framework in the Qedgix algorithm. When collecting information from adjacent nodes (other UAVs or users), the graph neural network can operate without being affected by the specific order of node feature inputs. This method underscores the capability of graph neural networks in processing dynamic and complex network, especially in flexibly adapting to collaborative tasks and decision-making processes within multi-UAV systems without the explicit need to consider node ordering, thus ensuring the model's adaptability to dynamic environmental changes.

\section{Experimental Results}

In this section, we conduct extensive experiments to evaluate the performance of the proposed Qedgix algorithm and compare it with existing methods. In the experimental setup, the task area is defined as a square of $1 \times 1 \, \text{km}^2$. The initial positions of the UAVs are all set at $[0.5,0.5]$, and the initial positions of the users are randomly distributed within the task area. All the UAVs are set to maintain a flight altitude of $H = 50 \, \text{m}$. The total number of time slots in these experiments is $K=80$. We propose a new unit, denoted by $\xi$ and defined as 40 meters, to enhance the accuracy of describing spatial relationships in experimental scenarios.  The operational speed of UAVs is set to $\xi / \text{s} $. The transmission range is consistently set to $3\xi$. The detection range is set to $7\xi$.

To ensure comparability and fairness of the experimental results, the Qedgix algorithm was configured with hyperparameters consistent with those of the benchmark algorithms used for comparison. This method minimizes potential biases introduced by varying parameter settings and provides a reliable basis for evaluating the performance of the Qedgix algorithm against established methods. Regarding action selection, an $\epsilon$ greedy strategy was adopted, with a 0.05 probability for random exploration. In all experiments involving the mixer network, mixing\_embed\_dim is consistently set to 32, and the hypernet\_embed is uniformly maintained  64. The experience memory has the capacity to store 5,000 transition samples, and the optimization is performed using the Adam optimizer with a learning rate of 0.005. The batch\_size during the algorithm training process is set to 128. In order to obtain an accurate estimation of algorithm performance, all results are obtained through multiple experiments.

% \begin{table}[ht]
%     \centering
%     \setlength{\tabcolsep}{3mm}{}
%     \caption{hyper-parameter setting of different algorithms}
%     \resizebox{0.9\columnwidth}{!}{
%     \begin{tabular}{c|cccc}
%     \toprule[0.5pt]\toprule[0.5pt]
%        Training Parameters  & QMIX & Qedgix & Agggnn & RgcnConv\\ \hline 
%        number of mixer layers   & 2 & 2 & 2 & 2 \\ \hline 
%        hidden dimension & 64 & 64 & 64 & 64 \\ \hline 
%        $lr$ & 5e-3 & 5e-3 & 5e-3 & 5e-3 \\ \hline 
%        optimizer  & Adam & Adam & Adam & Adam  \\ \hline 
%        t\_max & 1.5e+7 & 1.5e+7 & 1.5e+7 & 1.5e+7 \\ \hline 
%        gnn\_layers & $\backslash$ & 2 & 2 & 2 \\ \hline 
%        gnn\_type & $\backslash$ & EdgeConv & Agggnn & RgcnConv \\ \hline 
%        batch size & 128 & 128 & 128 & 128 \\\hline 
%        edge\_type & $\backslash$ & no & no & yes \\
%        \bottomrule[0.5pt]\bottomrule[0.5pt]
       
%     \end{tabular}
%     }
%     \vspace{-8pt}
%     \label{params}
% \end{table}

To comprehensively evaluate the performance of the Qedgix algorithm, this paper conducts detailed simulation comparisons between Qedgix and the original QMIX algorithm, as well as variants integrating different GNN models into the QMIX agent framework. These comparisons aims to reveal the advantages of the Qedgix algorithm in complex interactive environments.

\subsection{Compare the Performance of Different Algorithms}
% The architecture of the above algorithm mainly consists of two components: the agent model (Model) and the mixer network (Mixer). The role of the agent is to observe and collect environmental information and execute actions, providing local Q-value estimates for global decision-making. The function of the mixer network is to integrate the individual values of each agent to generate a global estimate. Based on this global estimate, each agent can make decisions according to this information.

% The architecture of the QMIX algorithm consists of two key parts, namely the agent model and the hybrid network (Mixer). The proposed Qedgix algorithm is based on the QMIX algorithm and integrates graph neural networks (GNN) to achieve a more effective information transfer mechanism. To comprehensively evaluate the performance of the Qedgix algorithm, the experimental design of this paper includes a detailed simulation comparison between the Qedgix algorithm and the original QMIX algorithm and variants of agents that fuse different GNN models into the QMIX algorithm. Through this comparison, we aim to accurately reveal the advantages and potential performance improvements of the Qedgix algorithm in handling complex interactive environments.

The architecture of the QMIX algorithm consists of two key components: the agent model and the mixer network. The agent model is responsible for the decision-making process of each agent, while the mixer network integrates the decision-making information of all agents to optimize the overall performance of multi-agent cooperation. Building upon the QMIX algorithm, this study introduces the Qedgix algorithm, which enhances the network structure of the agent model by incorporating GNN to facilitate a more effective information exchange mechanism.

% There are four models in this experiment, namely, Algorithm 1 $\left(model\colon RNN, mixer\colon QMIX\right)$, Algorithm 2 $\left(model\colon RNN+EdgeConv, mixer\colon QMIX\right)$, Algorithm 3 $\left(model\colon RNN+Agggnn, mixer\colon QMIX\right)$, Algorithm 4 $\left(model\colon RNN+RgcnConv, mixer\colon QMIX\right)$. More hyper-parameters of Qedgix and other compared algorithms are shown in Table \ref{params}.

Four algorithms are compared in this experiment:
\begin{enumerate}
    \item QMIX: $\text{model: RNN, mixer: QMIX}$
    \item Qedgix: $\text{model: RNN+EdgeConv, mixer: QMIX}$
    \item Algorithm 3: $\text{model: RNN+AggGNN, mixer: QMIX}$
    \item Algorithm 4: $\text{model: RNN+RGCN, mixer: QMIX}$
\end{enumerate}
AggGNN updates node states through aggregation operations, which gather information from neighboring nodes to facilitate the update. This method effectively captures local connectivity patterns between nodes, thereby enhancing the modeling capability of graph-structured data. Relational Graph Convolutional Network (RGCN) is specialized in handling graph data with multiple types of edge relationships. In RGCN, each relationship type has its own dedicated weight, making it suitable for modeling heterogeneous relationships between entities.

% AggGNN exemplifies the aggregation operation on nodes, updating node states by gathering information from neighboring nodes to capture local connectivity patterns among them. RGCN stands out in handling graphs with multiple types of relationships, assigning unique weights to each relationship type and making it apt for modeling the heterogeneous relations between entities. 

% More hyper-parameters of Qedgix and other compared algorithms are shown in Table \ref{params}.

% \begin{figure}[ht]
%   \centering
%   \includegraphics[width=0.95\columnwidth]{fig/newest/experiment_1.pdf}
%    % \vspace{-9pt}
%   \centering \caption{\added{Reward comparison for UAVs collecting user data with four different algorithms.}} 
%   \label{experiment1}
%    % \vspace{-9pt}
% \end{figure}

% \begin{figure}[ht]
%   \centering
%   \includegraphics[width=0.95\columnwidth]{fig/newest/Figure_9.pdf}
%    % \vspace{-9pt}
%   \centering \caption{\added{Reward comparison for UAVs collecting user data with four different algorithms.}} 
%   \label{experiment1}
%    % \vspace{-9pt}
% \end{figure}

\begin{figure}[ht]
  \centering
  \includegraphics[width=0.98\columnwidth]{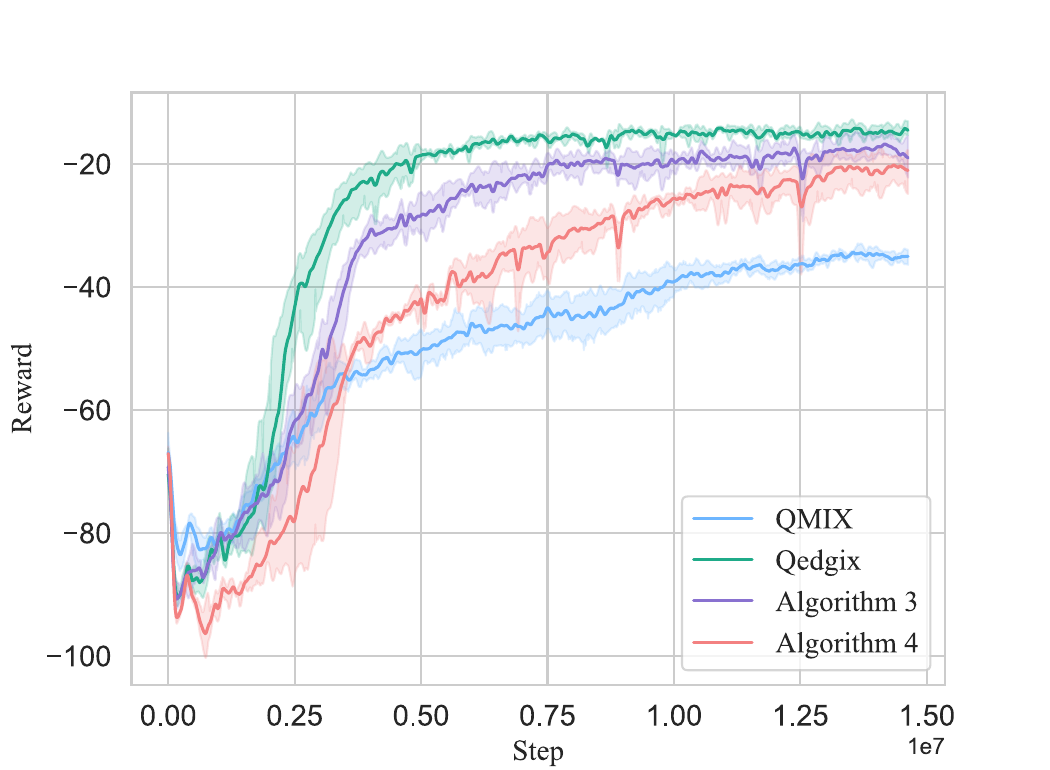}
   % \vspace{-9pt}
  \centering \caption{Reward comparison for UAVs collecting user data with four different algorithms.} 
  \label{experiment1}
   % \vspace{-9pt}
\end{figure}

\begin{figure*}[ht]
  \centering
  \includegraphics[width=1.90\columnwidth]{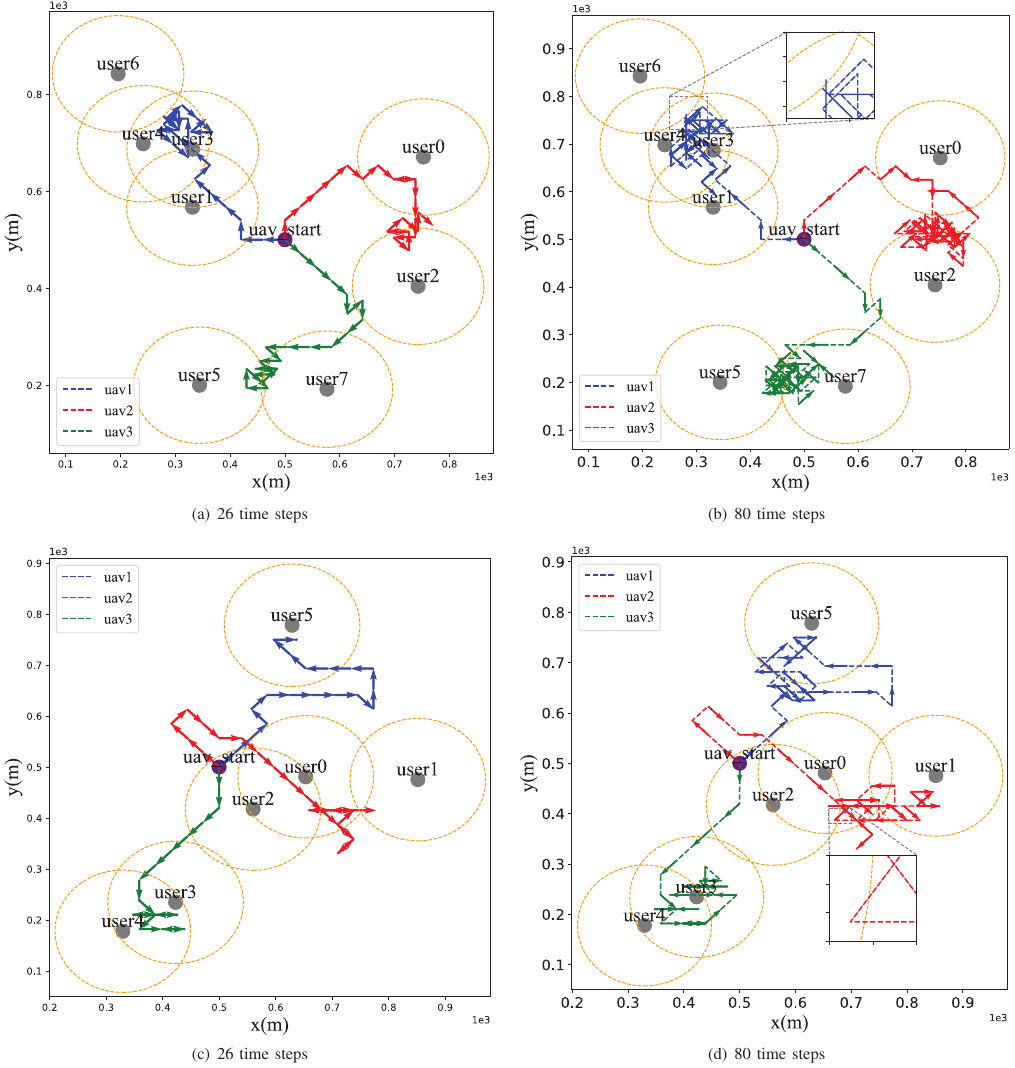}
  \centering \caption{Subfigures (a)-(b) show trajectories for the scenario with three UAVs and six users. Subfigures (c)-(d) show trajectories for the scenario with three UAVs and eight users.} 
  \label{fig5}
   % \vspace{-9pt}
\end{figure*}

% \begin{figure}[ht]
%   \centering
%   \includegraphics[width=1.90\columnwidth]{fig/newest/GNN_QMIX-9.png}
%    % \vspace{-9pt}
%   \centering \caption{Subfigures (a)-(b) show trajectories for the scenario with three UAVs and six users. Subfigures (c)-(d) show trajectories for the scenario with three UAVs and eight users.} 
%     \label{fig4:times}
%    % \vspace{-9pt}
% \end{figure}

Fig.~\ref{experiment1} presents the experimental results concerning the convergence of the proposed Qedgix algorithm alongside three comparative algorithms. Initially, the QMIX algorithm demonstrates a rapid convergence rate, likely due to its agents' relatively simple neural network architecture, which facilitates easier training in the early stages. However, as training progresses, Algorithms 3 and 4, which incorporate AggGNN and RGCN graph neural network structures, exhibit commendable convergence, albeit not reaching the performance level of Qedgix. This performance disparity stem from the fact that AggGNN and RGCN are primarily designed for static graphs and not align completely with the dynamic characteristics of temporal multi-agent control tasks. In contrast, the Qedgix algorithm shows an exceedingly swift convergence rate during training, significantly improving in terms of exploration efficiency and training efficacy, and displaying remarkable stability in performance, markedly surpassing the traditional QMIX algorithm and the other two GNN-enhanced algorithm variants. These findings strongly indicate the superiority of the Qedgix algorithm in handling complex reinforcement learning tasks, particularly in accelerating convergence speed and enhancing exploration capabilities.

\subsection{UAV Trajectories}
To investigate the trajectory optimization effect of the Qedgix algorithm for UAVs collecting user data, we visualized and analyzed the dynamic behavior of UAVs during the data collection process. Two sets of experiments are conducted using different combinations of UAVs and users: one set involves 3 UAVs with 6 users, and the other set involves 3 UAVs with 8 users. These setups are used to evaluate the effectiveness of the Qedgix algorithm in collecting user data.

% In Fig.~\ref{fig5}, each user is depicted as an entity with a yellow circle at its center, symbolizing the user's transmission range. Trajectories of UAVs at different time steps demonstrating their path planning. The arrows in Fig.~\ref{fig_sub41:time2} and \ref{fig_sub42:time2} are sparser than in Fig.~\ref{fig_sub41:time1} and \ref{fig_sub42:time1}, not due to increased UAV step size, but because an arrow is drawn every other step for clarity.

In Fig.~\ref{fig5}, each user is depicted as an entity with a yellow circle at its center, symbolizing the user's transmission range. Trajectories of UAVs at different time steps demonstrating their path planning. The arrows in Fig. 5(b) and 5(d) are sparser than in Fig. 5(a) and 5(c), not due to increased UAV step size, but because an arrow is drawn every other step for clarity.

Fig.~\ref{fig5} clearly demonstrates that three UAVs, initiating from the position [0.5,0.5], search for users in three distinct directions, benefiting the UAVs in maximizing the detection of scattered users in the area. From Fig. 5(a)--5(b), UAV2 initially heads northwest and, upon detecting no users, reverses its course to the southeast for data collection, identifying user0, user1, and user2 as targets en route, subsequently conducting data collection among these users. Similarly, after UAV3 designates user3 and user4 in the southwest as targets, it conducts a return trip between the two users for data collection. UAV1, after a search period in the northeast and discovering only user5, immediately begins the data collection process. As demonstrated in Fig. 5(c)--5(d), UAV1 heads towards the northwest, UAV2 towards the northeast, and UAV3 towards the southeast. UAV1 detects user1, user3, user4, and user6 in the northwest direction and designates them as its data collection targets. In the northeast direction, UAV2 identifies the data collection targets user0 and user2. After flying for some time, UAV3 detects the presence of UAV2 in its northeast direction, changes its course to fly westward, and subsequently identifies the data collection targets user5 and user7.

The zoomed-in subfigure in Fig. 5(b) shows that UAV1 collects data at the boundary of user6's transmission range and immediately turns back, avoiding excessive dwelling and thus maintaining trajectory efficiency. Similarly, the zoomed-in subfigure in Fig. 5(d) demonstrates good performance, where UAV2 collects data at the boundary of user2's transmission range and then turns back to collect information from other users in the area. Therefore, it is evident that the Qedgix algorithm demonstrates high performance by providing reasonable action instructions for UAVs. UAV1, UAV2, and UAV3 can rapidly locate areas where user groups congregate and then execute back-and-forth maneuvers to continuously collect user data.

\subsection{The Impact of UAV Detection Range}

Next, we investigate the performance disparities of the Qedgix algorithm under different environmental conditions in this subsection. We compare it with two other baseline algorithms, i.e., the QMIX algorithm and Algorithm 3. To assess the adaptability of Qedgix algorithm to changes in UAV detection range sizes, we have designed a series of experiments where the detection range of UAVs incrementally expands from $4\xi$ to $9\xi$. Throughout these experiments, the number of UAVs and users remains constant, with 3 UAVs and 5 users, allowing for an accurate evaluation of how variations in UAV detection ranges affect the performance of the algorithm.

% \begin{figure}[ht]
%   \centering
%   \includegraphics[width=0.9\columnwidth]{fig/newest/Figure_6.pdf}
%    % \vspace{-9pt}
%   \centering \caption{\added{Mean AoI values vs. UAV detection range for three algorithms.}} 
%   \label{experiment3}
%    % \vspace{-9pt}
% \end{figure}

\begin{figure}[ht]
  \centering
  \includegraphics[width=0.9\columnwidth]{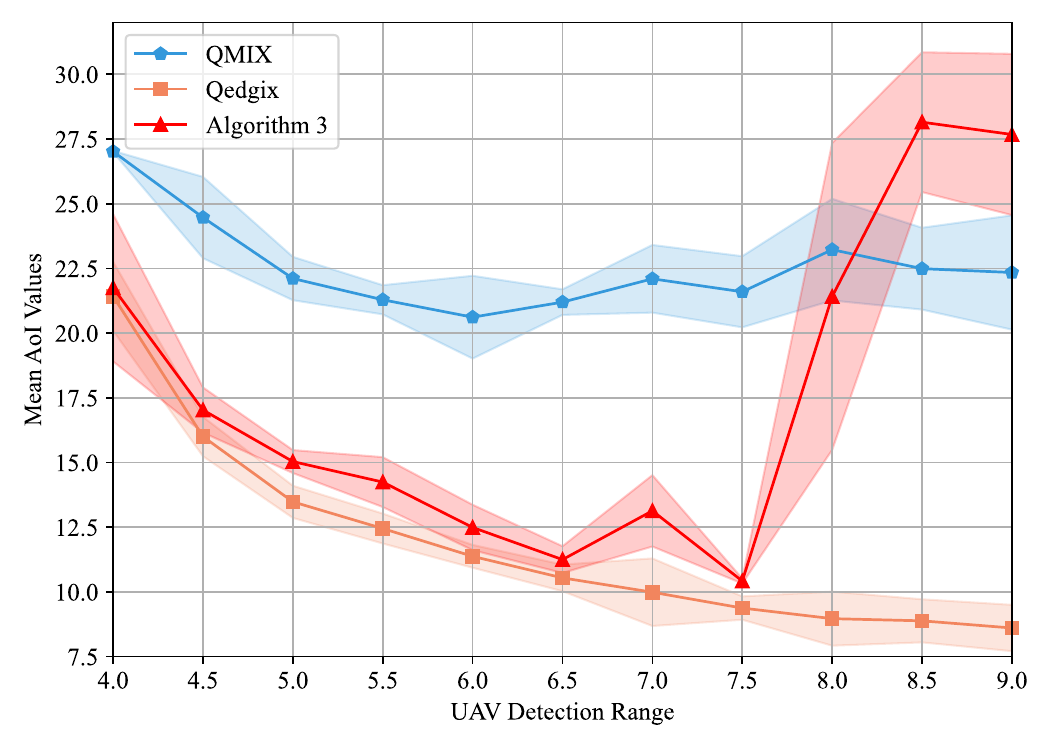}
   % \vspace{-9pt}
  \centering \caption{Mean AoI values vs. UAV detection range for three algorithms.} 
  \label{experiment3}
   % \vspace{-9pt}
\end{figure}

% As can be seen from the Figure 4, with the change of detection range, the Qedgix algorithm exhibits better results in training UAV swarms, effectively realizes UAV trajectory planning, and minimizes AoI values. In addition, the Qedgix algorithm exhibits excellent stability. As the detection range expands, the UAV swarm exhibits a stronger "global vision", thus more effectively coordinating the completion of tasks. In contrast, the stability of the QMIX algorithm is weak, and its performance does not improve significantly with the increase of detection range.

Fig.~\ref{experiment3} illustrates that, as the UAV detection range increases, the Qedgix algorithm exhibits superior performance in training UAVs, effectively optimizing their trajectory planning and significantly reducing the average AoI. Particularly, as the detection capabilities of UAVs gradually enhance, the Qedgix algorithm can implicitly provide the UAVs with an expanded field of view, maximizing their spatial awareness. This enhanced perspective enables a more effective collaborative operation to accomplish tasks. Additionally, the Qedgix algorithm demonstrates excellent stability, with the increase in UAV detection range leading to a stable decrease in the mean AoI values of the data collected within the area. In contrast, the QMIX algorithm displays limitations in performance, as its performance does not improve proportionally with the increase in detection range. Although Algorithm 3 approaches Qedgix algorithm in performance, we can observe that when the UAV detection range is $4\xi$ to $6.5\xi$, there is still a performance gap compared to the Qedgix algorithm. Additionally, as the detection range increases, the performance of Algorithm 3 drops sharply. This is due to the limitation of AggGNN in handling graph data with dynamic topological changes. These experimental outcomes not only demonstrate the substantial potential of Qedgix algorithm in UAVs control and data collection but also showcase its ability to maintain efficient operations under varying environmental conditions when changes occur in the UAV's detection range.

\subsection{The Impact of Different Agent Scale}

Finally, we evaluate the performance of the Qedgix algorithm in managing UAVs and user groups of varying sizes. For comparison, we also use the QMIX algorithm as a benchmark. The number of UAVs varies from 2 to 4, and the user group sizes range from 4 to 8. This setup is designed to simulate service demands of different densities.

\begin{figure}[ht]
  \centering
  \includegraphics[width=0.9\columnwidth]{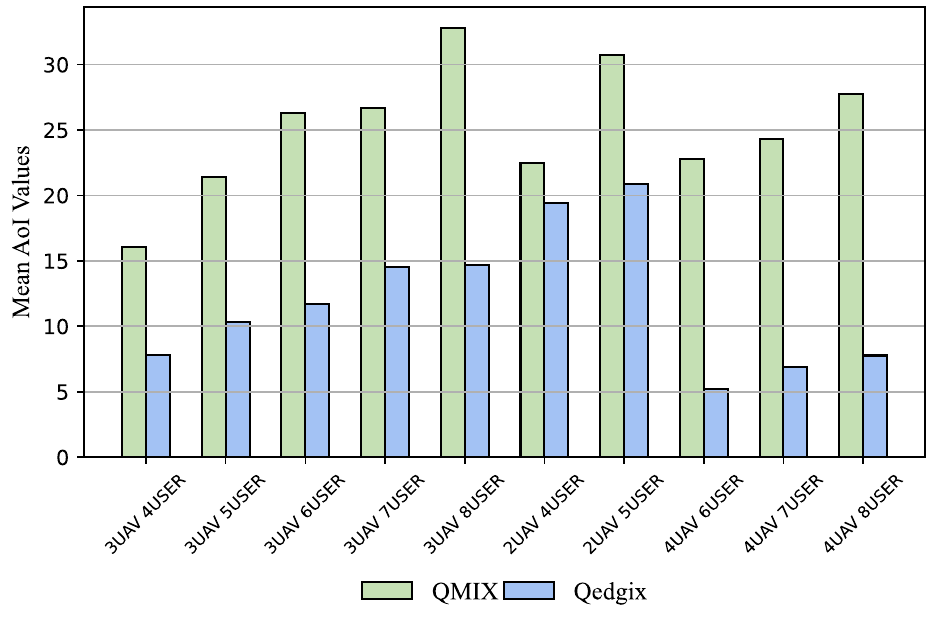}
   \vspace{-9pt}
  \centering \caption{Comparison of mean AoI values across different UAV and user configurations for QMIX and Qedgix algorithms..} 
  \label{experiment4}
   \vspace{-9pt}
\end{figure}

% Based on the analysis of the above experimental results, it can be clearly observed in figure 5 that the average AoI collected by the Qedgix algorithm is smaller when the agent scale in the environment changes. This shows that the algorithm can guide the training of drone swarms more efficiently, thus realizing UAV trajectory planning more effectively.

% Fig.~\ref{experiment4} provides a comparative analysis of the mean AoI values between the Qedgix and QMIX algorithms with varying scales of agents. 

Fig.~\ref{experiment4} compares the mean AoI values of the Qedgix and QMIX algorithms across varying scales of agents. The figure reveals that the Qedgix algorithm consistently maintains a lower mean AoI values across different settings, particularly noticeable in scenarios with a larger number of UAVs and users. For instance, in the 2 UAVs and 4 users configuration, Qedgix demonstrates a marginal performance advantage over QMIX, likely due to the sparse distribution of UAVs and users which impedes the full potential of the GNN in facilitating tight interconnections and information dissemination. Conversely, in the configuration with 4 UAVs and 8 users, the performance of the Qedgix algorithm significantly outperforms that of the QMIX algorithm, attributable to the integration of GNN in the agent network structure, enabling efficient message passing for information processing in dense relational networks. These findings underscore the adaptability and efficiency of the Qedgix algorithm in managing UAVs when dealing with users of different numbers in the environment. Despite the varying numbers of UAVs and users across different scenarios, Qedgix effectively optimizes UAVs trajectory planning and ensures the timeliness of data collection.

\section{Conclusion and Future Work}
% In this paper, we have investigated a multi-agent reinforcement learning (MARL) algorithm for unmanned aerial vehicle (UAV) planning, targeting the optimization of scenarios where UAVs function as mobile access nodes for user data collection. By integrating EdgeConv with the QMIX framework, we proposed the Qedgix algorithm to minimize the mean Age of Information (AoI) values of user data through the optimization of UAV trajectories. We modeled UAVs and users as dynamic nodes and performed hidden layer message passing, significantly enhancing the policy network's ability to perceive complex environmental information. Experimental results demonstrated that the Qedgix algorithm, incorporating graph neural networks, excels in handling complex environmental information and shows superior performance in practical application scenarios. The application of this technology to UAV networks can reduce the computational power requirements for centralized controllers through its distributed optimization characteristics, thereby improving the trajectory planning performance of UAV networks based solely on observable states in unknown environments. Future work will consider the impact of variations in user data packet sizes on the data collection process and aim to optimize UAV energy efficiency, further enhancing the algorithm's overall performance and practical value.
In this paper, we investigate a MARL algorithm for UAV trajectory planning, targeting the optimization of scenarios where UAVs function as mobile access nodes for user data collection. By integrating EdgeConv with the QMIX framework, we propose the Qedgix algorithm to minimize the mean AoI values of user data by optimizing UAV trajectories. We model UAVs and users as dynamic nodes and perform hidden layer message passing, significantly enhancing the policy network's ability to perceive complex environmental information. Experimental results demonstrate that the Qedgix algorithm, incorporating graph neural networks, excels in handling complex environmental information and shows superior performance in practical application scenarios. Our approach can reduce the computational power requirements for centralized controllers through its distributed optimization characteristics, thereby improving the trajectory planning performance of UAV networks based solely on observable states in unknown environments. Future work will consider the impact of variations in user data packet sizes on the data collection process and aim to optimize UAV energy efficiency, further enhancing the algorithm's overall performance and practical value. Meanwhile, since recent study report the intrinsic gradient-based weakness of neural networks \cite{zhang2024quantum}, we will also study the resilience of our model in terms of such vulnerabilities.

\bibliography{ref}

\bibliographystyle{IEEEtran}

\ifCLASSOPTIONcaptionsoff
  \newpage
\fi

\end{document}